\documentclass[prc,twocolumn,showpacs,amsmath,amssymb,superscriptaddress,floatfix,nofootinbib]{revtex4}
\usepackage{mathrsfs,bm}
\usepackage{longtable,lscape}
\usepackage{txfonts}
\usepackage{amssymb}
\usepackage{indentfirst}
\usepackage{graphicx,,booktabs}
\usepackage{multirow}
\usepackage{color}
\usepackage{amssymb}

\begin{document}
\title{Gottfried sum rule from  maximum entropy method quark distributions with DGLAP evolution and DGLAP evolution with GLR-MQ-ZRS corrections}

\author{Chengdong Han}
\affiliation{Institute of Modern Physics,Chinese Academy of Sciences, Lanzhou 730000, China}
\affiliation{University of Chinese Academy of Sciences, Beijing 100049, China}

\author{Qiang Fu}
\affiliation{Institute of Modern Physics,Chinese Academy of Sciences, Lanzhou 730000, China}
\affiliation{University of Chinese Academy of Sciences, Beijing 100049, China}
\affiliation{Lanzhou University, Lanzhou 730000, China}

\author{Xurong Chen}
\email{xchen@impcas.ac.cn} \affiliation{Institute of Modern Physics,Chinese Academy of Sciences, Lanzhou 730000, China}

\begin{abstract}
\textbf{Abstract}: A new method to test the valence quark distribution of nucleon obtained from the maximum entropy method using Gottfried sum rule by performing DGLAP equations with the GLR-MQ-ZRS corrections and original LO~/~NLO DGLAP equations are outlined. The test relies on a knowledge of the unpolarized electron-proton structure function $F_2^{ep}$ and electron-neutron structure function $F_2^{en}$ and the assumption that Bjorken scaling is satisfied. In this work, the original Gottfried summation value obtained by the integrals of the structure function at different Q$^{2}$ is in accordance with the theoretical value 1/3 under the premise of light-quark flavour symmetry of nucleon sea, whether it is the result from the dynamics evolution equations or the result from global QCD fits of PDFs. Finally, we present the summation value of the LO / NLO DGLAP global fits of PDFs under the premise of light-quark flavour asymmetry of nucleon sea. According to analysis the original Gottfried summation value with two evolution equations at different $Q^{2}$, we can know that the valence quark distributions of nucleon obtained by the maximum entropy method are effective and reliable.
\end{abstract}


\maketitle

\section{Introduction}
\label{SecI}
Up to now, there exists a number of sum rules for unpolarized and polarized structure functions, some of which are rigorous results and other which rely on more or less well justified assumptions~\cite{1}.
The Adler sum rule \cite{2} is exact and has no Quantum chromodynamics~(QCD) perturbative corrections, but its experimental verification is at a very low level of accuracy \cite{3}. This constant 2 of Adler sum rule is the result of the local commutation relations of the time components of the hadronic weak current \cite{4}, which is based on the fundamental quark structure of the standard model.
By contrast, the corresponding Gottfried sum rule \cite{5} for charged lepton scattering was based merely on valence quarks picture and is modified both by perturbative and by non-perturbative effects \cite{6,7}.
The original Gottfried sum rule states that the integral over Bjorken variable x of a difference of electron-proton and electron-neutron structure functions is a constant 1/3 under flavour symmetry in the nucleon sea ($\bar{u}(x)=\bar{d}(x)$), which is independent of the negative four-momentum transfer squared $Q^{2}$. Some experimental results were achieved from electron, and muon \cite{8} scattering on isoscalar targets or on Hydrogen target \cite{9} deep-inelastic
scattering (DIS). For non-singlet Mellin moment neutrino and charged-lepton DIS, with the N = 1 moments corresponding to the Adler and Gottfried sum rules \cite{5,6,7}.

In this paper, we test valence quark distributions of nucleon obtained from the Maximum Entropy Method (MEM) by original Gottfried sum rule using DGLAP equation \cite{10} with GLR-MQ-ZRS corrections~(DGLAP-GLR-MQ-ZRS equations~/~IMParton16 package)~\cite{11} at different $Q^2$, which is compared with the results of the original DGLAP evolution equations and the latest global fits of parton distribution functions.
The most important correction to DGLAP evolution equations is to consider the parton-parton recombination effect. The IMParton16 package, we developed the dynamical parton model about the origin of parton distributions, and extended the initial evolution scale down to $Q^2$ $\sim$ 0.1 GeV$^{2}$.
For the leading-order (LO) and next-to-leading-order (NLO) DGLAP equations evolution, we use the modified Mellin transformation method by CANDIA \cite{12} to calculate original Gottfried summation value under the premise of light-quark flavour asymmetry of the nucleon sea ($\bar{u}(x)= \bar{d}(x)$). The starting scale for the LO and NLO evolution is Q$^{2}$ = 1 GeV$^{2}$. Finally, we give the summation value of the LO / NLO DGLAP latest global fits of parton distribution functions under the premise of light-quark flavour asymmetry of the nucleon sea ($\bar{u}(x)\neq\bar{d}(x)$). And the obtained summation value at different Q$^{2}$ are nearly consistent with experimental observations.

The organization of the paper is as follows. A non-perturbative initial input of valence quark distributions of nucleon by MEM are introduced in Section II. Section III discusses Gottfried sum rule. Section IV presents comparisons of DGLAP-GLR-MQ-ZRS results with LO / NLO DGLAP equations results under the premise of light-quark flavour symmetry in the nucleon sea, as well as calculate the summation value of LO / NLO DGLAP latest global fits of parton distribution functions under the premise of light-quark flavour asymmetry in the nucleon sea. Finally, a summary is given in Section \ref{SecV}.

\section{A non-perturbative initial input from the quark-parton model}
\label{SecIV}
The quark model is a classification scheme for hadrons in terms of their valence quarks and assumes that baryons are composed of three quarks and mesons of a quark and an anti-quark.
The solutions of the QCD evolution equations for parton distributions of the nucleon at high $Q^2$ depend on the initial parton distributions at low starting scale $Q^2_0$. According to the Quark Model, an ideal assumption is that the nucleon consists of only three valence quarks at extremely low $Q_0^2$. Hence, a non-perturbative initial input of the nucleon includes merely three valence quarks, which is the simplest input of nucleon \cite{13}. In the dynamical parton distribution functions model, all sea quarks and gluons are QCD radioactively generated from valence quarks at high scale $Q^2$.
The simple function form to approximate valence quark distribution is the time-honored canonical parametrization $f(x) = A x^B (1-x)^C$ \cite{14}.
Thus, the simplest parametrization of the naive non-perturbative input of the proton by MEM \cite{15} is written as
\begin{equation}
\begin{aligned}
&u_v^p(x,Q_0^2)=7.191 x^{0.286}(1-x)^{1.359},\\
&d_v^p(x,Q_0^2)=13.068 x^{0.681}(1-x)^{3.026}.
\end{aligned}
\label{Parametrization}
\end{equation}
In addition, the valence quark distributions of the free neutron by previous work~\cite{16} is written as
\begin{equation}
\begin{aligned}
&u_v^n(x,Q_0^2)=16.579 x^{0.780}(1-x)^{3.267},\\
&d_v^n(x,Q_0^2)=8.678 x^{0.369}(1-x)^{1.511}.
\end{aligned}
\label{Parametrization}
\end{equation}

By performing DGLAP-GLR-MQ-ZRS evolution equations \cite{11}, one can determine valence quark distributions of nucleon at high $Q^{2}$ with the initial non-perturbative input obtained by MEM \cite{15,16}.
We get the specific low starting scale $Q_0^2=0.0671$ GeV$^2$ for the naive non-perturbative input, by performing QCD evolution for the second moments of the valence quark distributions
\cite{17} and the measured moments of the valence quark distributions at a higher $Q^2$ \cite{18}.
The running coupling constant $\alpha_s$ for the leading order and the current quark masses are the parameters of perturbative QCD involved in the evolution equations \cite{11,15}.
For the LO and NLO DGLAP equations evolution, we use the modified Mellin transformation method by CANDIA \cite{12}. The starting scale for the LO and NLO evolution is Q$^{2}$ = 1 GeV$^{2}$.

\section{Gottfried sum rule}

In the proton there are two up valence quarks ($u_v$) and one down valence quark ($d_v$).
In fact, each quark distribution function q$_{i}$(x)~(i = u, d, s) always contains the sum of two parts, including the valence quark $q_{i}^{(v)}$ and the sea quark $q_{i}^{(s)}$ distribution function.
\begin{equation}
\begin{aligned}
&q_i(x)=q_i^{(v)}(x) + q_i^{(s)}(x).\\
\end{aligned}
\label{InitialValence}
\end{equation}

According to the definition of the distribution functions, the integrals of all distribution functions (the quarks and anti-quarks distribution functions q$_{i}$(x) and $\bar{q}$$_{i}$(x)) within the proton should give the valence quark number. Therefore the valence sum rules for the non-perturbative inputs are as follows
\begin{equation}
\begin{aligned}
&\int_0^1 [u(x)-\bar{u}(x)]dx=2,\\
&\int_0^1 [d(x)-\bar{d}(x)]dx=1.\\
&\int_0^1 [s(x)-\bar{s}(x)]dx=0.
\end{aligned}
\label{ValenceSum}
\end{equation}
Through the transformation of Eq.~(4), one can get that
\begin{equation}
\begin{aligned}
&\int_0^1 [\frac{2}{3}(u(x)-\bar{u}(x))-\frac{1}{3}(d(x)-\bar{d}(x))]dx=1,
\end{aligned}
\label{ValenceSum}
\end{equation}
\begin{equation}
\begin{aligned}
&\int_0^1 [\frac{2}{3}(d(x)-\bar{d}(x))-\frac{1}{3}(u(x)-\bar{u}(x))]dx=0.
\end{aligned}
\label{ValenceSum}
\end{equation}
The Eq.~(5) corresponds to the proton with the charge of 1. The Eq.~(6) correspond to neutron with the charge of 0. The reason is that proton and neutron are isospin doublet, and up and down quarks are also isospin doublet. So the distribution of up quark in neutron should be the same as that of down quark in proton.

According to the Quark-Parton model, the structure function of nucleon is written as
\begin{equation}
\begin{aligned}
&2xF_1(x)=F_2(x)=\sum\limits_i e_i^2 xf_i(x),\\
\end{aligned}
\label{InitialValence}
\end{equation}
which is called the Callan-Gross expression \cite{19}. Where i is the flavor index, $e_{i}$ is the electrical charge of the quark of flavour i (in the unites of the electron charge), and $xf_{i}$ is the momentum fraction of the quark with flavor i. The structure functions of proton and neutron obtained by the deep inelastic scattering of the charged lepton on protons and neutrons are respectively,
\begin{equation}
\begin{aligned}
&\frac{1}{x}F_2^{ep}(x)=\frac{4}{9}[u(x)+\bar{u}(x)]+\frac{1}{9}[d(x)+\bar{d}(x)]+\frac{1}{9}[s(x)+\bar{s}(x)]\\
\end{aligned}
\label{InitialValence}
\end{equation}
\begin{equation}
\begin{aligned}
&\frac{1}{x}F_2^{en}(x)=\frac{4}{9}[d(x)+\bar{d}(x)]+\frac{1}{9}[u(x)+\bar{u}(x)]+\frac{1}{9}[s(x)+\bar{s}(x)]\\
\end{aligned}
\label{InitialValence}
\end{equation}

For the proton, it can set as:
\begin{equation}
\begin{aligned}
&s_v(x)=\bar{s}_{v}(x)=\bar{u}_{v}(x)=\bar{d}_{v}(x)=0,\\
&u_s(x)=\bar{u}_{s}(x)=d_{s}(x)=\bar{d}_{s}(x)=s_{s}(x)=\bar{s}_{s}(x)=\frac{1}{6}S(x),\\
&S(x)={u}_{s}(x)+{d}_{s}(x)+\bar{u}_{s}(x)+\bar{d}_{s}(x)+{s}_{s}(x)+\bar{s}_{s}(x).
\end{aligned}
\label{InitialValence}
\end{equation}
The S(x) is the sea quarks sum of proton, neglecting the heavy quark's sea quark wave function.

By bringing the constraints Eqs. (3) and (10) into the Eqs. (8) and (9), then one can get that
\begin{equation}
\begin{aligned}
&f(x)=\frac{1}{x}(F_2^{ep}(x)-F_2^{en}(x))=\frac{1}{3}(u_v(x)-d_v(x)).\\
\end{aligned}
\label{InitialValence}
\end{equation}
Where the f(x) is a function of the Bjorken scaling variables x.

From Eq.~(11), one can know that the difference between the proton structure function $F_2^{ep}$ and the neutron structure function $F_2^{en}$ comes only from the contribution of the valence quarks, and the contribution of the sea quarks just be canceled with each other. Thus, the measurement of the proton and neutron structure functions will provide information about valence quarks. The integral of Eq.~(11) with the constraints of Eq.~(3) and Eq.~(4) is as follows
\begin{equation}
\begin{aligned}
&I=\int_0^1 \frac{dx}{x}(F_2^{ep}(x)-F_2^{en}(x)),\\
\end{aligned}
\label{InitialValence}
\end{equation}
Where I is the integral summation value of Eq. (12). Theoretically this integral value is constant 1/3, which is called the original Gottfried sum rule \cite{9} under flavour symmetric of nucleon sea. In this paper, we use I$_{i}$(Q$^2$) to represent the original Gottfried summation value from two evolution equations (DGLAP-GLR-MQ-ZRS equations and DGLAP equations) at different Q$^2$.

Gottfried studied high-energy electron-neucleon scattering, meson-nucleon reactions and the spectroscopy of heavy-quark bound states. Then he proposed the Gottfried sum rule \cite{5,6,9} for deep inelastic scattering to test the elementary quark model.
The corresponding Gottfried sum rule for charge lepton-nucleon DIS was involving a form factor of nucleon. Within the quark-parton model, the corresponding isospin sum rule in the case of charged-lepton-nucleon DIS is as follows:

\begin{equation}
\begin{aligned}
&I=\int_0^1 \frac{dx}{x}(F_2^{lp}(x)-F_2^{ln}(x))\\
 &=\int_0^1 dx[\frac{1}{3}(u_{v}(x)-d_{v}(x))+\frac{2}{3}(\bar{u}(x)-\bar{d}(x))]\\
 &=\frac{1}{3}-\frac{2}{3}\int_0^1dx(\bar{d}(x)-\bar{u}(x)).
\end{aligned}
\label{InitialValence}
\end{equation}
If the neucleon sea were flavour symmetric, with $\bar{u}(x)=\bar{d}(x)$, one should have $I_{G}(Q^{2})$= 1/3.
If the nucleon sea were flavour asymmetric, namely $\bar{u}(x)\neq\bar{d}(x)$, one should have $I_{G}(Q^{2})\neq$1/3. Moreover, this result is supported by the existing neutrino-nucleon DIS data \cite{3}, and the most detailed analysis of muon-nucleon DIS data of NMC Collaboration \cite{8}. It is worth noting that there are also some other works \cite{20} of the light-quark flavour asymmetry deviation from the canonical value 1/3 for the Gottfired sum rule.

\section{Results and discussion}

The DGLAP equations describe the evolution of quark and gluon densities with Q$^{2}$, which is based on the parton model and perturbative QCD theory. The DGLAP-GLR-MQ-ZRS evolution equation is based on DGLAP equation, which is mainly consider the parton recombination effect. The theoretical work of the parton recombination effect was first proposed by  by Gribov, Levin and Ryskin (GLR) \cite{21}, then Mueller and Qiu (MQ) put forward the recombination probabilities for gluons to go into gluons or into quarks in a low-density limit \cite{22} and give a detailed calculation. Furthermore, Zhu, Ruan and Shen (ZRS) further present a set of new and concrete evolution equations about parton recombination corrections \cite{23}.

It is worth noting that the number density of parton increases rapidly in the small x area.
In a small x, the number density of parton increases to a certain extent so that the quanta of partons overlap spatially.
Therefore the parton-parton recombination effect becomes essential at small x area, which is can effectively prevent the continuous increase of the cross sections near their unitarity limit.

In fact, the GLR-MQ-ZRS corrections can be very effective in slowing down the parton splitting at low scale Q$^{2}$ $<$ 1~GeV$^{2}$. Up to now, ZRS have considered all the recombination functions for gluon-gluon, quark-gluon and quark-quark processes \cite{23}. Due to the gluon density is obviously greater than the quark density at small x, the gluon-gluon recombination effect is dominant in calculation \cite{11}. Therefore, we use the simplified form of the DGLAP equations with GLR-MQ-ZRS corrections (DGLAP-GLR-MQ-ZRS equations) in the analysis \cite{11}.

In order to accurately test the validity of the DGLAP equation with GLR-MQ-ZRS corrections about parton distribution function evolution at different $Q^{2}$, we perform the integral of Eq.~(12) which is completely independent of $Q^2$ in theory. By applying the DGLAP-GLR-MQ-ZRS evolution equations, the quark distribution functions of proton and neutron ( Eqs. (1) and (2)) are evolved to high $Q^{2}$, and the structure functions of proton and neutron $F^{p,n}_{2}$ under different $Q^2$ are further calculated.

\begin{figure}[htp]
\begin{center}
\includegraphics[width=0.45\textwidth]{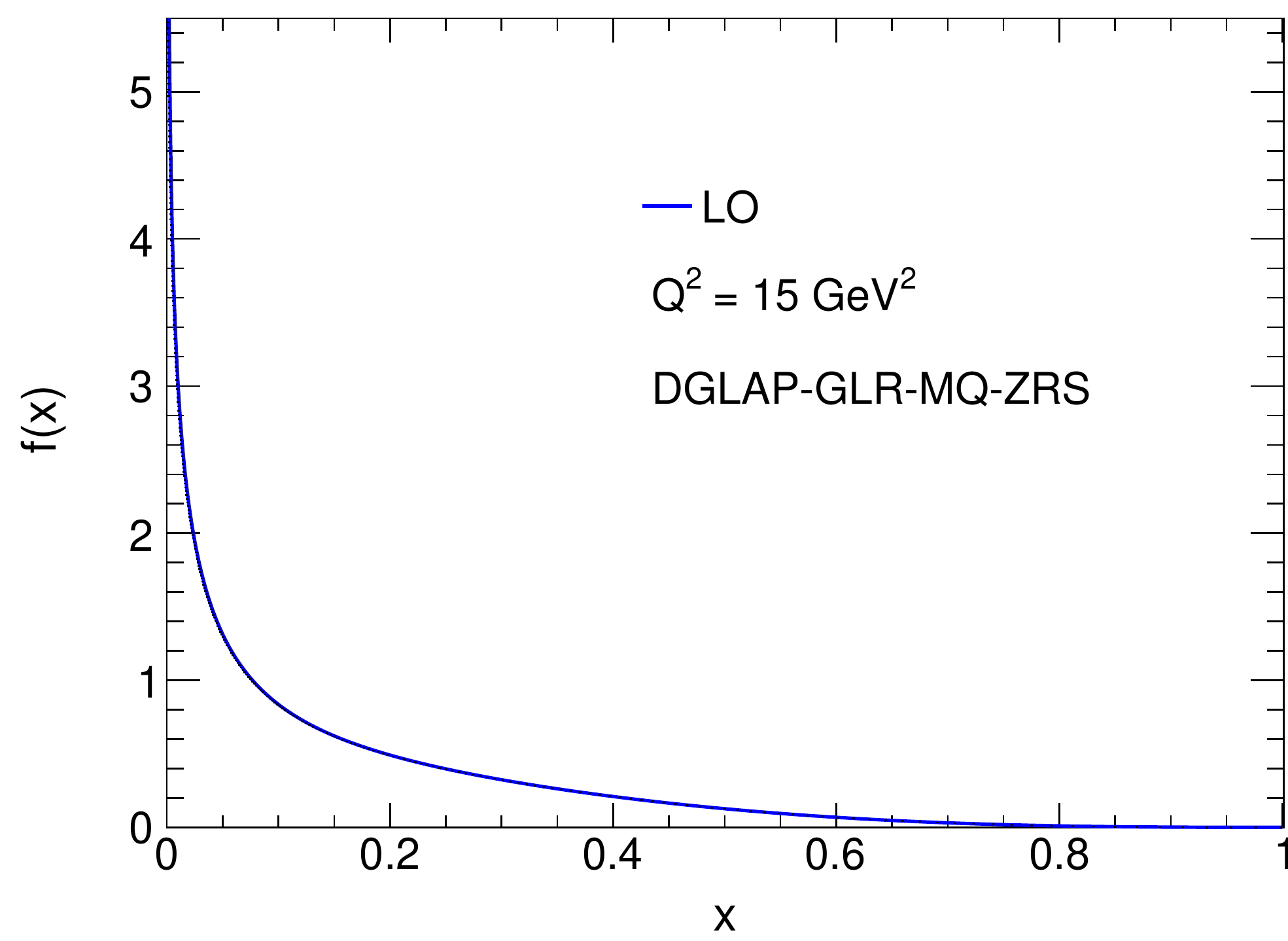}
\caption{(Color online)~
f(x) as a function of the Bjorken scaling variables x.
}
\label{Entropy}
\end{center}
\end{figure}

We take the DGLAP-GLR-MQ-ZRS equations as dynamical evolution equation to obtain the distribution of the right end of the Eq.~(11). The integral value of the area below the curve in Figure 1 is 0.3333 at $Q^{2}$ = 15 GeV$^{2}$. It's obvious that the result from the DGLAP-GLR-MQ-ZRS equations is in good agreement with the theoretical values 1/3 under light-quark flavour symmetry.

After that, we take the DGLAP equations as dynamical evolution equation to obtain the distribution of the right end of the Eq.~(11), the starting scale Q$^{2}_{0}$ = 1 GeV$^{2}$ for the LO and NLO evolution with naive non-perturbative input, which is from the modified Mellin transformation method by CANDIA \cite{12}. By applying the DGLAP evolution equations, the quark distribution functions of proton and neutron from MEM ( Eqs. (1) and (2)) as initial input) are evolved to high $Q^{2}$. Then one can get the original Gottfried summation value I$_{i}(Q^2)$ of LO and NLO with light-quark flavour symmetry.

\begin{figure}[htp]
\begin{center}
\includegraphics[width=0.45\textwidth]{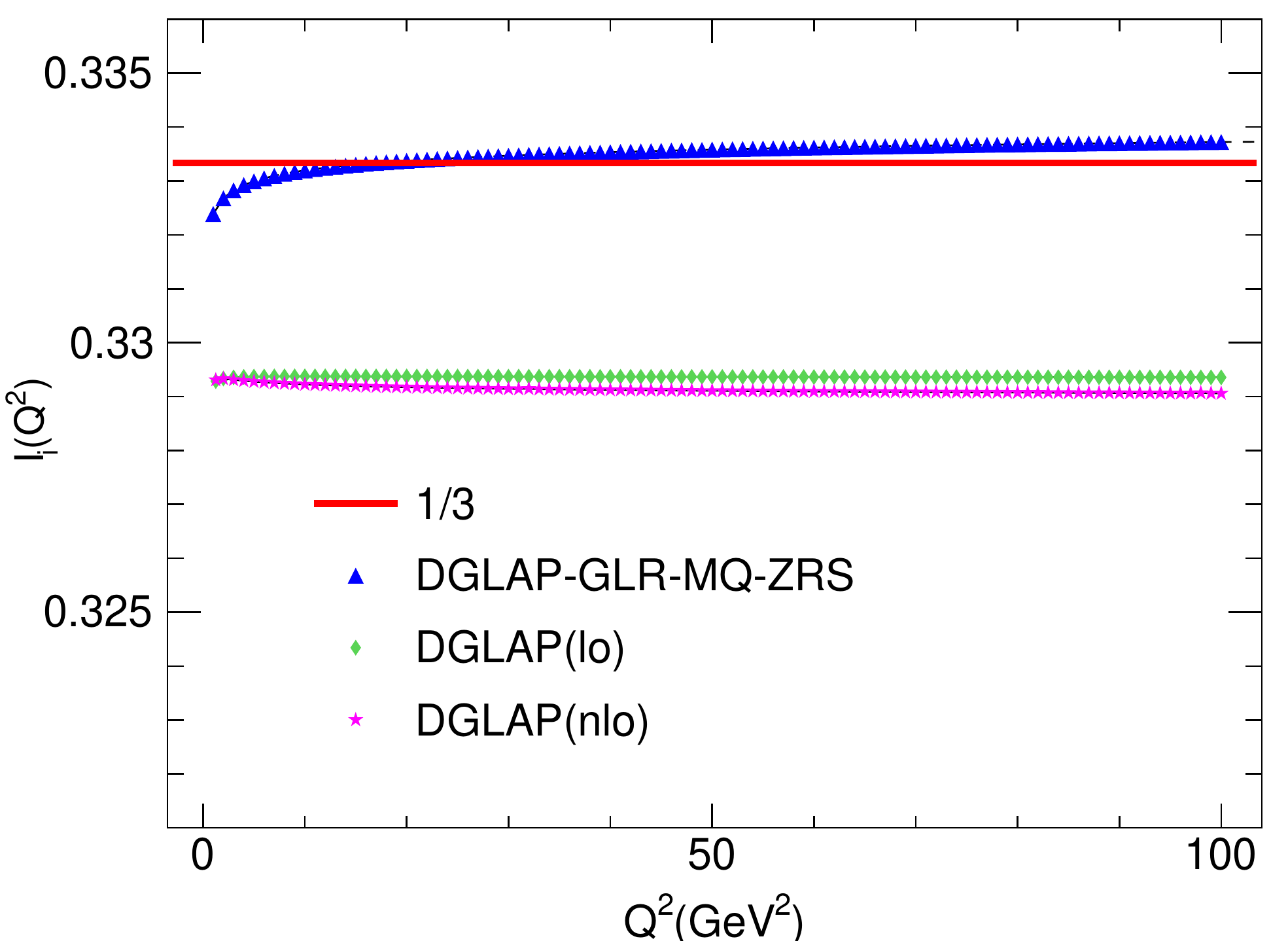}
\caption{(Color online)~
Comparisons of original Gottfried summation value from DGLAP-GLR-MQ-ZRS equations (triangle) with results from DGLAP equations LO (rhombus) and NLO (star) at different $Q^{2}$ under the premise of light-quark flavour symmetry $\bar{u}(x)=\bar{d}(x)$ .
}
\label{Entropy}
\end{center}
\end{figure}

Figure 2 shows the comparisons of the original Gottfried summation value I$_{i}(Q^2)$ from DGLAP-GLR-MQ-ZRS equations with results from DGLAP equations with LO and NLO at different $Q^{2}$. The red solid line in figure 2 represents the theoretical value 1/3. Triangle, rhombus and star represent the original Gottfried summation value at different $Q^{2}$ given by the DGLAP-GLR-MQ-ZRS evolution equations, and DGLAP evolution equations with LO and NLO, respectively.

It is apparent that the original Gottfried summation value of DGLAP-GLR-MQ-ZRS equations have a smaller deviation than the summation value of LO / NLO DGLAP equations. But when you get down to the detail, one can find that summation value from DGLAP equations are not exactly equivalent to the theoretical value of 1/3, but slightly smaller than 1/3. Moreover, the summation value from the NLO DGLAP equations is slightly smaller than the summation value of the LO DGLAP equations evoluation, which is from $\alpha_{s}^{2}$-level perturbative QCD correction. These corrections compared with the experimental analysis results turn out to be small and cannot be responsible for the significant discrepancy between experimental results and the naive expectation of 1/3.
It is noteworthy that original quark-parton model expression for the original Gottfried sum rule is modified by perturbative QCD contributions when nucleon sea were flavour symmetric in Ref. \cite{7}.
\begin{figure}[htp]
\begin{center}
\includegraphics[width=0.45\textwidth]{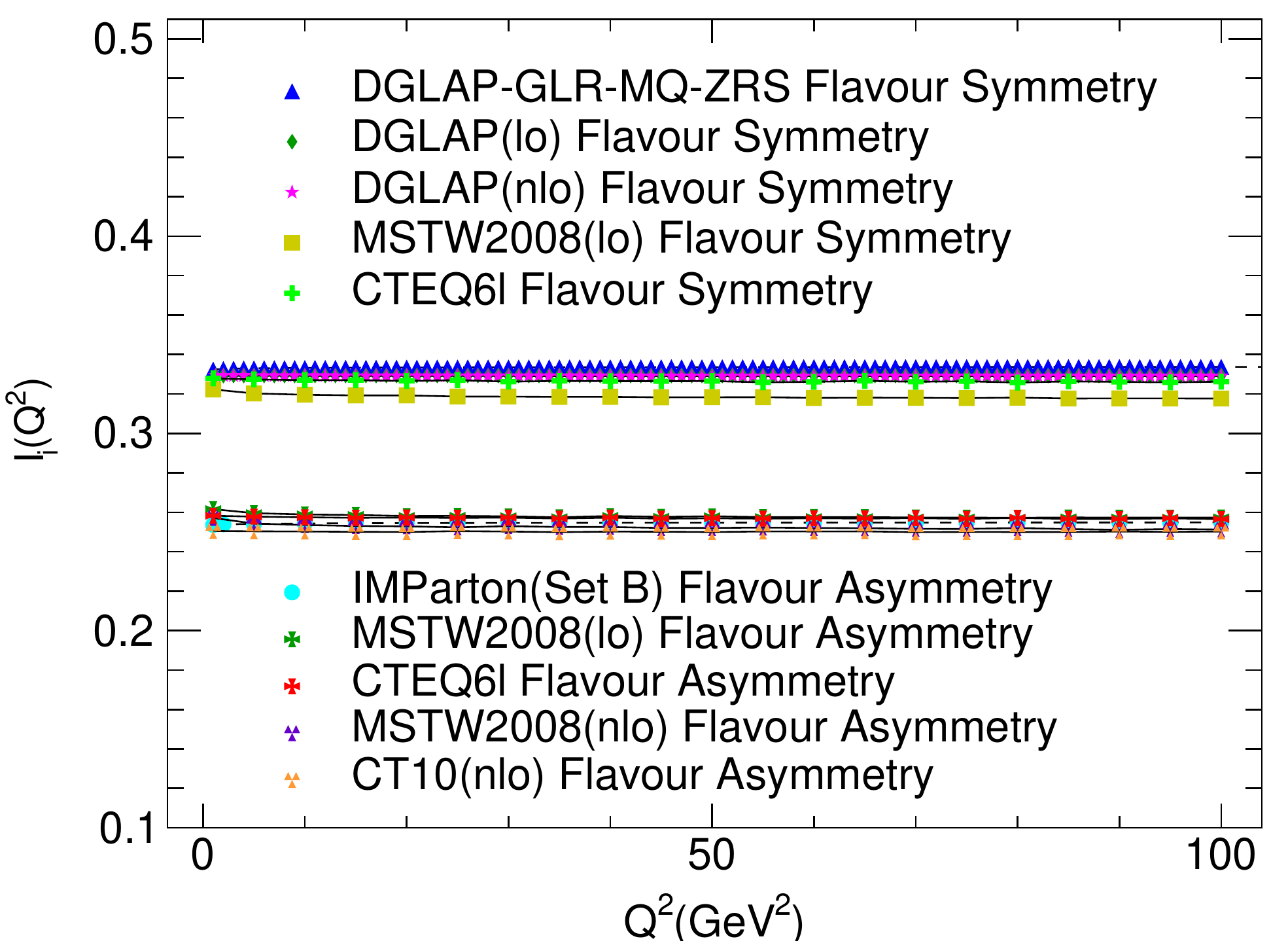}
\caption{(Color online)~
Comparisons of summation value from two dynamics evolution equations (DGLAP-GLR-MQ-ZRS / DGLAP equations) with global QCD fit from IMParton16 (Set B), MSTW(LO / NLO) and CTEQ6l / CT10(NLO) at different $Q^{2}$ under the premise of light-quark flavour symmetry or asymmetry of the nucleon sea.
}
\label{Entropy}
\end{center}
\end{figure}

In order to more intuitively analyze the summation value under the light-quark flavour symmetry and asymmetry of the nucleon sea, we compare the results from the dynamical evolution equations with the global fits from IMParton16 (Set B), MSTW (LO /NLO)~\cite{24} and CTEQ6l \cite{25} / CT10(NLO)~\cite{26}, as shown in Figure 3.

From the Figure 3, the results of the two evolution equations (DGLAP-GLR-MQ-ZRS / DGLAP equations) are basically consistent with the results with
global QCD fit MSTW08(LO) (square), CTEQ6l (cross) on the premise of flavour symmetric, and are approximately equal to the theoretical value of 1/3.
In the previous analysis, the summation value under flavour symmetry of the nucleon sea cannot describe the violation from experimental analysis data. What we have to admit that this result from the quark-parton flavour-symmetric prediction is very naive.

By analyzing the results of the light-quark flavour symmetry, we can know that it does not affect the necessity to introduce flavour-asymmetry ($\bar{u}(x)\neq\bar{d}(x)$) for the description of experimental analysis results for the Gottfried sum rule.
Figure 3 show the direct results from IMParton16(Set B), MSTW(LO / NLO) and CTEQ61 / CT10(NLO) PDFs by computing Eq. (12) with the use of Eq. (7). It is clearly indicates that the violation of the theoretical 1/3 with the light-quark asymmetry of the nucleon sea, which is in agreement with the experimental analysis results. What's more, one can find that the summation value from NLO DGLAP global fits MSTW(NLO) and CT10(NLO) PDFs by computing Eq. (12) with the use of Eq. (7) are slightly smaller than the LO DGLAP global fits MSTW(LO) and CTEQ6l PDFs, which is from the perturbative QCD correction.

\section{summary}\label{SecV}

In this work, the valence quark distribution function of nucleon at low Q$_{0}^{2}$ obtained by the MEM is used as non-perturbative initial input.
Then the parton distributions of the nucleon are evaluated dynamically at high $Q^{2}$ by the DGLAP-GLR-MQ-ZRS equations and LO~/~NLO DGLAP equations.
Then we get the unpolarized electromagnetic structure functions for proton and neutron $F^{p,n}_{2}$. Through the calculation of Eq. 12, one can further obtain the Gottfried summation value.

This is an interesting attempt to test the valence quark distribution function of nucleon obtained by the MEM via Gottfried sum rule by performing DGLAP-GLR-MQ-ZRS equations and DGLAP equations. The original Gottfried summation value obtained by Eq. (11) with different $Q^2$ is in accordance with the theoretical value 1/3 under the light-quark flavour symmetry of nucleon sea.
It is apparent that the original Gottfried summation value of DGLAP-GLR-MQ-ZRS equations have a smaller deviation than the summation value of LO / NLO DGLAP equations.
Moreover, the summation value from the NLO DGLAP equations is slightly smaller than the summation value of the leading order DGLAP equations evoluation, which is from $\alpha_{s}^{2}$-level perturbative QCD correction. The correction is small.
It should be mentioned that the naive theoretical summation value equal to 1/3 is very preliminary compared with the existing experimental analysis results.
Finally, we give the summation value from Figure 3, which is not equal to 1/3 with NLO DGLAP evoluation and the global fits from IMParton16 (Set B), MSTW (LO / NLO)~\cite{24},CTEQ6l~\cite{25} / CT10(NLO)~\cite{26} under light-quark flavour asymmetry. This is the necessity of introducing light-quark flavour asymmetry in the nucleon sea for the description of experimental analysis results.

The Gottfired sum rule verifies the reliability of non-perturbative initial input of valence quark distributions from the starting low scale $Q^{2}_0$ by performing DGLAP-GLR-MQ-ZRS equations. The DGLAP-GLR-MQ-ZRS equations based on DGLAP equations with parton-parton recombination corrections is an important innovation, which shows that the nonlinear effects of parton-parton recombination are non-negligible at low Q$^{2}$.
According to the results of the above analysis, the valence quark distribution functions of nucleon obtained by the MEM as initial input are valid and reliable.

\noindent{\bf Acknowledgments}:
The authors would like to thank Rong Wang for helpful and fruitful suggestions.
This work is supported by the National Basic Research Program of China (973 Program) 2014CB845406.


\begin{thebibliography}{90}

\bibitem{1}
J. Soffer, arXiv:hep-ph/0409333.

\bibitem{2}
Adler S L. Phys. Rev., {\bf 143}, 1144 (1966).

\bibitem{3}
P.C. Bosetti et al, Nucl. Phys. B {\bf 142}, 1 (1978); \\
J.C.H. deGroot et al. Z. Phys. C Particles and Fields {\bf 1}, 143 (1979); \\
S.M. Heagy et al., Phys. Rev. D {\bf 23}, 1045 (1981); \\
M. Jonker et al., Phys. Lett. 109 {\bf B}, 133 (1981); \\
P.C. Bosetti et al., Nucl. Phys. B {\bf 203}, 362 (1982); \\
Bergsma et al., Phys. Lett. B {\bf 123}, 269 (1983); \\
H. Abramowicz et al., Z. Phys. C - Particles and Fields {\bf 17}, 283 (1983); \\
H. Abrarnowicz et al., Z. Phys. C - Particles and Fields {\bf 25}, 29 (1984); \\
D.B. MacFarlane et al., Z. Phys. C - Particles and Fields {\bf 26}, 1 (1984); \\
WA25 Collaboration, D. Allasia et al., Z. Phys C {\bf 28}, 321 (1985).

\bibitem{4}
Stephen L. Adler, arXiv:0905.2923.

\bibitem{5}
K. Gottfried, Phys. Rev. Lett.  {\bf 18} 1174 (1967).

\bibitem{6}
D. J. Broadhurst, A. L. Kataev and C. J.Maxwell, Phys. Lett. B {\bf 590}, 76 (2004).

\bibitem{7}
A. L. Kataev and G. Parente, Phys. Lett. B {\bf 566}, 120 (2003).

\bibitem{8}
A. Bodek et al., Phys. Rev. D {\bf 20}, 1471 (1979); \\
D. Bollini et al., Phys. Lett. B {\bf 104}, 403 (1981); \\
J.J. Aubert et al., Phys. Lett. B {\bf 105}, 322 (1981); \\
A.R. Clark et al., Phys. Rev. Lett. {\bf 51}, 1826 (1983); \\
M. Arneodo, et al., New Muon Collaboration, Phys. Rev. D {\bf 50} 1 (1994);\\
A.L. Kataev, arXiv:hep-ph/0311091 (2003).

\bibitem{9}
S.J. Wimpenny: In Proc. Int. Conf. on High Energy Physics, Brighton, 1983;\\
J.J. Aubert et al., Phys. Lett. B {\bf 123}, 123 (1983).

\bibitem{10}
Y. L. Dokshitzer, Sov. Phys. JETP {\bf 46}, 641 (1977); \\
V. N. Gribov and L. N. Lipatov, Sov. J. Nucl. Phys. {\bf 15}, 438 (1972); \\
G. Altarelli and G. Parisi, Nucl. Phys. B {\bf 126}, 298 (1977).

\bibitem{11}
X. Chen, J. Ruan, R. Wang, W. Zhu, and P. Zhang, Int. J. Mod. Phys. E {\bf 23}, 1450057 (2014); \\
R. Wang, X. Chen and Q. Fu, Nucl. Phys. B {\bf 920}, 1 (2017); \\
Rong Wang and Xurong Chen, Chin. Phys. C {\bf 41}, 053103 (2017), https://github.com/lukeronger/IMParton; \\
Wei Zhu, Rong Wang, Jianhong Ruan, Xurong Chen, Pengming Zhang, Eur. Phys. J. Plus {\bf 131}, 6 (2016).

\bibitem{12}
Alessandro Cafarella, Claudio Corian and Marco Guzzi, Comput. Phys. Comm. 179 665 (2008);\\
A.D. Martin, et al., Eur. Phys. J. C 23 73 (2002);\\
A.D. Martin, et al., Phys. Lett. B 531 216 (2002).

\bibitem{13}
G. Parisi and R. Petronzio, Phys. Lett. B {\bf 62}, 331 (1976); \\
V. A. Novikov, M. A. Shifman, A. I. Vainshtein, and V. I. Zakharov, JETP Lett. {\bf 24}, 341 (1976); \\
M. Gl\"uck and E. Reya, Nucl. Phys. B {\bf 130}, 76  (1977); \\
X. Chen, J. Ruan, R. Wang, W. Zhu, and P. Zhang, Int. J. Mod. Phys. E {\bf 23}, 1450057 (2014).

\bibitem{14}
Jonathan Pumplin, Daniel Robert Stump, Joey Huston, Hung-Liang Lai,
Pavel Nadolsky, and Wu-Ki Tung, J. High Energy Phys. {\bf 07}, 012 (2002).

\bibitem{15}
Rong Wang, Xurong Chen, Phys. Rev. D {\bf 91}, 054026 (2015).

\bibitem{16}
Chengdong Han, Jiangshan Lan, Qiang Fu and Xurong Chen, arXiv:1801.01387.

\bibitem{17}
E. Reya, Phys. Rep. {\bf 69}, 195 (1981).

\bibitem{18}
M. Gl\"uck, E. Reya, and A. Vogt, Eur. Phys. J. C {\bf 5}, 461 (1998).

\bibitem{19}
C. G. Callan, D. J. Gross, Phys. Rev. Lett. {\bf 22}, 156 (1969).

\bibitem{20}
S. Kumano, Phys. Rep. 303 183 (1998);\\
G.T. Garvey, J.C. Peng, Prog. Part. Nucl. Phys. 47 203 (2001);\\
M. Karliner, H.J. Lipkin, Phys. Lett. B 533 60 (2002).

\bibitem{21}
L. V. Gribov, E. M. Levin and M. G. Ryskin, Phys. Rep. {\bf 100}, 1 (1983).

\bibitem{22}
A. H. Mueller and Jianwei Qiu, Nucl. Phys. B {\bf 268}, 427 (1986).

\bibitem{23}
Wei Zhu, Nucl. Phys. B {\bf 551}, 245 (1999); \\
Wei Zhu and Jianhong Ruan, Nucl. Phys. B {\bf 559}, 378 (1999); \\
Wei Zhu and Zhenqi Shen, High Energy Physics and Nuclear Physics {\bf 29}, 109 (2005).

\bibitem{24}
A.D. Martin, W.J. Stirling, R.S. Thorne and G. Watt, Eur. Phys. J. C 63 189(2009).

\bibitem{25}
Pavel M. Nadolsky et al., Phys. Rev. D {\bf 78}, 013004 (2008).

\bibitem{26}
H.-L. Lai, M. Guzzi, J. Huston, Z. Li, P. M. Nadolsky, J.
Pumplin, and C.-P. Yuan, Phys. Rev. D {\bf 82}, 074024 (2010)

\end{thebibliography}
\end{document}